**Direct observations of chiral spin textures in van der Waals magnet $Fe_3GeTe_2$ nanolayers**

Hong Wang[1,2]#, Cuixiang Wang[1,3]#, Yan Zhu[4]#, Zi-An Li[1]*, Hongbin Zhang[5], Huanfang Tian[1], Youguo Shi[1,3], Huaixin Yang[1,3], Jianqi Li[1,3,5]*

[1]Beijing National Laboratory for Condensed Matter Physics, Institute of Physics, Chinese Academy of Sciences, Beijing 100190, China

[2]College of Material Science and Opto-Electronic Technology, University of Chinese Academy of Sciences, Beijing 100049, China

[3]School of Physical Sciences, University of Chinese Academy of Sciences, Beijing 100049, China

[4]College of Science, Nanjing University of Aeronautics and Astronautics, Nanjing 210016, China

[5]Institute of Materials Science, Technical University of Darmstadt, Darmstadt 64287, Germany

[6]Songshan Lake Materials Laboratory, Dongguan, Guangdong 523808, China

# These authors contributed equally to this work.

*Correspondence should be addressed to Z.-A. Li (zali79@iphy.ac.cn) and J.Q. Li (ljq@iphy.ac.cn)

## Abstract

In two-dimensional van der Waals (vdW) magnets, the presence of magnetic orders, strong spin-orbit coupling and asymmetry at interfaces is the key ingredient for hosting chiral spin textures. However, experimental evidences for chiral magnetism in vdW magnets remain elusive. Here we demonstrate unambiguously the formation of chiral spin textures in thin $Fe_3GeTe_2$ nanoflakes using advanced magnetic electron microscopy and first-principles calculations. Specifically, electron holography analyses reveal the spin configurations of Néel-type, zero-field-stabilized skyrmions in 20-nm-thick $Fe_3GeTe_2$ nanoflakes at cryogenic temperature. *In situ* Lorentz transmission electron microscopy measurements further provide detailed magnetic phase diagrams of chiral spin textures including spirals and skyrmions in $Fe_3GeTe_2$ as a function of temperature, applied magnetic field and specimen thickness. First-principles calculations unveil a finite interfacial Dzyaloshinskii-Moriya interaction in the Te/$Fe_3Ge$/Te slabs that induces the spin chirality in $Fe_3GeTe_2$. Our discovery of spin chirality in the prototypical vdW $Fe_3GeTe_2$ opens up new opportunities for studying chiral magnetism in two-dimensional vdW magnets from both fundamental and applied perspectives.

## Introduction

The past years have seen a fast development of many new concepts and physical phenomena in two-dimensional (2D) van der Waals (vdW) magnets including the intrinsic magnetic orders at the 2D atomic limit[1,2] and the novel functionalities of hybrid vdW heterostructures for spintronics applications[3,4]. $Fe_3GeTe_2$ (FGT), a prototypical vdW magnet, has demonstrated a rich variety of physical phenomena. For example, bulk FGT exhibits itinerant ferromagnetism with a relatively high Curie temperature ($T_C$ = ~230 K) and a strong out-of-plane magnetic anisotropy[5,6]. When thinned down to the 2D limit, FGT atomic layers can sustain ferromagnetic ordering[7,8]; remarkably, the $T_C$ for FGT atomic layers can be enhanced either by further reducing dimensions via nanopatterning[9] or by electrical ion-gating[10]. Electronic band theories and experiments revealed the effects of strong electron correlations[11], the Kondo lattice physics[12], and the topological nodal lines physics[13] in bulk FGT. Electrical transport measurements discovered very large anomalous Hall currents[13] and topological Hall effect in FGT atomic layers[14]. These novel properties found in FGT have motivated many efforts looking into their specific spin textures. For instance, magnetic force microscopy (MFM) imaging has revealed complex magnetic domain structures including magnetic labyrinthine patterns[8,9], circular dots within magnetic branches[15], and singly, doubly-walled bubbles of Néel-type domain walls[16]. Néel-type spin textures observed in the perpendicularly magnetized FGT suggest the existence of spin chirality enabled by the Dzyaloshinskii-Moriya interaction[17–19], but contradicted by symmetry argument that FGT belongs to a centrosymmetric space group of $P6_3mmc$. Therefore, experimental evidences (if exists[20], the underlying physics) for the existence of chiral spin textures in vdW FGT remain elusive.

To investigate complex spin textures in vdW magnets, the capability of real-space magnetic imaging with high spatial resolution and high magnetic sensitivity is essential. Compared with optical and x-ray magnetic microscopy of broad-beam nature or scanning probe microscopy with only surface spin sensitivity, transmission electron microscopy (TEM) based techniques like Lorentz microscopy (LM)[21] and off-axis electron holography (EH)[22] offer a spatial resolution below 5 nm and high magnetic sensitivity. LM and EH have been successfully applied to study various complex noncollinear spin textures in different systems[23–25]. Here we employed the LM and EH to directly image the spin textures in thin FGT nanoflakes exfoliated from the bulks. Quantitative magnetic phase images from EH measurements allow us to assess the characteristics of chiral Néel-type skyrmions in the perpendicularly magnetized FGT nanoflakes. In situ LM observations with varying specimen temperature and external magnetic fields provide a detailed temperature-field (T-H) magnetic phase diagram and specimen-thickness-dependent chiral spin textures in FGT thin films. Furthermore, our first-principles calculations reveal a finite interfacial Dzyaloshinskii-Moriya interaction residing at the Te/Fe and Fe/Te interfaces being responsible for the occurrence of chiral magnetism in FGT.

## Results and Discussion

As depicted schematically in Fig. 1a, the prominent crystallographic features in the hexagonal structure[5,6](space group of $P6_3mmc$) of $Fe_3GeTe_2$ are the trilayer of Fe/FeGe/Fe sandwiched by Te layers and the ~3 Å gap of Te-Te vdW layer separating each trilayer. Such structural features are clearly resolved in the electron diffraction and high-resolution TEM images of thin FGT nanoflakes. Figure 1b shows two high-angle-annual-dark-field (HAADF) images of [001]- and [120]-oriented FGT nanoflakes, together with the corresponding electron diffraction patterns. These atomic-resolved images and the sharp diffraction spots reveal the FGT samples are of high-quality and single crystalline.

Figure 1c shows the temperature-dependent magnetization (M-T) measured along *c*-axis and *ab*-plane using zero-field-cool (ZFC) and field-cool (FC) protocols. The Curie temperature $T_C$ of the FGT is measured to be ~ 230 K from these M-T curves, in consistent with previously reported values for stoichiometric $Fe_3GeTe_2$ single crystals[6,26]. The magnetization dips in the temperature range of 160 - 200 K and a fast reduction of magnetization below 160 K are clearly seen in the ZFC M-T curves along *c*-axis and *ab*-plane. The magnetization dips were further confirmed by AC magnetic susceptibility measurements; Figure 1d marks the temperatures at which two peaks appear in the real part of the AC susceptibility. Our magnetization curves are in consistent with the previous reports[26], in which the magnetization reduction was associated with a ferromagnetic to antiferromagnetic transition. In the present study, however, we demonstrate that the magnetization reduction and dips are the consequence of the formations of Néel-type chiral spin spirals in thick FGT layers, as seen in the magnetic images below.

To directly image the magnetic domain structures in FGT nanoflakes, we carried out EH and LM measurements. In the electron-wave description, fast electrons acquire a phase change, $\phi(r)$, when traversing a thin specimen. Thus, retrieval of $\phi(r)$ by EH or LM provides direct accesses to the electric and magnetic properties of the specimen[21,22]. It is noted that the basis of magnetic contrast in EH or LM requires a component of magnetization curl parallel to the electron beam direction[27]. When imaged with normal incident electrons, Néel-type domain walls or skyrmions produce no magnetic contrast because of the lack of such component of magnetic curl[28,29]. Néel spin textures are usually made visible by titling specimen to produce the required component of magnetization curl[28,29]; it also constitutes a practical way in the LM to distinguish between Néel-, Bloch- and hybrid-Néel-Bloch-type spin textures[30]. Figure 2a (upper panel) present a spin model of Néel skyrmion that consist of an extended out-of-plane spin domain bounded by a narrow circular wall of radially outward pointing spins. The Néel skyrmion is tilted by 20-degree, and the corresponding simulated magnetic phase and Lorentz images are shown in the lower panels in Fig. 2a, displaying the characteristic contrast features of Néel-type skyrmion.

Figure 2b shows a typical LM image of 20-nm-thick FGT nanoflake at 115 K under zero-applied-field, in which the prominent features of circular dots can be well recognized as Néel-type skyrmions because such contrast features can only be seen when the thin specimen was tilted by some degree and under out-of-focus conditions.

Series of specimen tilting and LM defocus-imaging experiments were carried out to confirm the existence of Néel-type skyrmions, as seen in the Supplementary Fig. S2. It is evident that the skyrmions exhibit a size distribution, and Fig. 2c presents a histogram of skyrmions diameters with a mean value of 82 nm and a standard deviation of 25 nm. The considerable small sizes of skyrmions of sub-100 nm found in vdW FGT are comparable to that in a limited number of magnetic multilayers including Ir/Co/Pt[31] and Co/Pd[29]. Because of the presence of size distribution, skyrmions in thin FGT nanoflakes do not form a regular lattice. The reproducible formations of skyrmions in FGT nanoflakes were further examined by a repetition of zero-field cyclic cooling-heating processes. We observed that the locations for the skyrmions are different for each experiment; that is, skyrmions are not pinned at specific crystal sites that may contain crystalline defects. The effect of cooling rate on the skyrmion formation was studied by varying cooling rate of the specimen from 1 to 10 K/min. For these cooling rates, no obvious differences in terms of skyrmion formation process and the size distribution were found. The size distribution observed in FGT may be associated with the multi-stability of skyrmions of different sizes found in magnetic multilayers[32].

To reveal the internal spin configurations of these circular dots, we carried out EH measurements that allow one to quantitatively analyze the spin texture via its projected magnetic phase images. Figure 2d shows the total phase image that comprises of electrical and magnetic phases, reconstructing from electron holograms taken at 115 K and under zero-field conditions. Clearly, the circular dots exhibit a prominent white-and-dark contrast feature, closely resembling the simulated magnetic phase of 20-degree-tilted Néel skyrmions in Fig. 2(a). Since the mechanically exfoliated FGT nanoflakes are very flat (see thickness measurement via energy-filtered TEM in the Supplementary Fig. S1), the electric phase contributions can be well approximated as an added phase offset of a constant value; thus, the phase variations can be approximated as magnetic phase[25]. Figure 2e shows a line profile extracted from an area marked in Fig. 2d, displaying the magnetic phase profile across a tilted skyrmion.

We now proceed to the systematic LM observations of magnetic evolution of chiral spin textures in FGT nanoflakes in response to variations in temperature and magnetic field (see videos v1 and v2 in the Supplementary Materials). Figure 3a and 3b present the temperature-field (T-H) phase diagram of a 20-nm-thick FGT nanoflake, summarizing the magnetic transition between different states driven by temperature and externally applied magnetic field. The specimens were first cooled from room temperature through Curie temperature ($T_C$ of ~230 K) to 10 K (lowest temperature attainable by liquid helium holder) under zero-field condition using the ZFC protocol. Above ~150 K, no visible contrast of magnetic structure was observed in the LM images; FGT nanoflakes were likely in a single domain state in the temperature range of 150 – 230 K. Upon further decreasing temperature below 150 K, strong contrast variations occurred in the form of fluctuating magnetic spirals and then rapidly condensed into individual circular dots that can be identified as chiral Néel skyrmions. The skyrmions are found to remain stable down to 10 K under zero-field conditions, with little changes in their locations and sizes. We also carried out series of cyclic cooling experiments to verify the repeatability (at the same specimen area) and the reproducibility (in different areas of new specimens with similar thickness) of the formations of zero-field skyrmions in FGT nanoflakes.

Next, we investigate the effects of external magnetic field on the stability of zero-field skyrmions by applying field $H_{\perp}$ normal to FGT nanoflakes. The videos in the Supplementary Materials recorded the detailed magnetic skyrmion evolution under varying $H_{\perp}$. Briefly, with increasing $H_{\perp}$, zero-field stabilized skyrmions were observed to shrink first under low fields, small-sized skyrmions collapsed under moderate fields, and eventually under high fields all large-sized skyrmions were saturated to a field-polarized ferromagnetic (FM) state. The saturation fields as a function of temperature were determined, which mark the boundaries between the skyrmionic and the FM states in Fig. 3a. Lastly, we reduced the applied fields $H_{\perp}$ from above saturation back to zero-field, and recorded the resulting magnetic changes (video v1 in the Supplementary Materials). Upon decreasing $H_{\perp}$ to certain values, magnetic spirals emerged from the FM state. Moreover, only in a relatively high temperature range from ~110 to 150 K the FM-to-spiral transition can be observed. Outside this temperature range, no formation of spirals (or skyrmions) was seen and the specimen remained in the field-polarized FM state. We note that although the spirals and the skyrmions occupied the specific T-H region of the phase diagram (Fig. 3a and 3b), their formations highly depend on the magnetizing protocols: in-field skyrmions persist upon increasing field, whereas in-field spirals emerge upon decreasing field. Hence, we conclude that skyrmions in FGT nanoflakes are thermodynamic stable phase and spirals metastable phase. The absence of spiral formation below 110 K and the fact that no observation of spiral-to-skyrmion transition at 150 K for a long period of time (tens of minutes for LM observations) suggest there exists a significant energy barrier between these two states.

Figure 3c shows the melting of zero-field skyrmions by heating process, as also seen in the video v2 of the Supplementary Materials. Upon heating in the temperature range from 10 to 170 K, the skyrmions appeared to only weakly alter their positions and sizes. They started to undergo strong alteration when the specimen temperature was raised above 170 K. In Fig. 3(c), skyrmions progressively dissolved into spirals at 179 K, and the spirals constantly fluctuated until they disappeared above 189 K.

Figure 3d presents the thickness-dependent chiral spin textures in FGT nanoflakes, which was deduced from LM observations for a number of specimens of various thickness. It is evident that only a narrow thickness range (around 15 – 25 nm; 20 to 30 trilayers of $Fe_3Ge$) supports the formation of zero-field Néel skyrmions. For thicker specimens (25 - 35 nm), short segments of spirals and skyrmions coexist; with even larger thickness (> 35 nm), e.g., 55 nm in Fig. 3d, the magnetic spirals dominate. The spirals in thicker specimens were found to transform into skyrmions by applying magnetic field normal to the specimen plane (Fig. S4 in the Supplementary Materials). When the applied field was switched off, only magnetic spirals emerged, exhibiting similarly magnetizing behavior as the zero-field skyrmion in thin FGT nanoflake. For thinner specimens (<15 nm), neither skyrmions nor spirals were observed under zero-applied-field. The absence of magnetic domain structures in few-layered FGT nanoflakes can be attributed to the strong dimensionality effects; that is, the considerable reduction of Curie temperature and anisotropies for few-layered FGT. [7,10]

The thickness-dependent skyrmion-to-spiral transition under zero-field can be used to estimate the strength of interfacial DMI in FGT. According to the effective medium method[30,32] for magnetic multilayers and the equation $\sqrt{8A(Q-1)\mu_0 M_s^2} - \pi D_i = \mu_0 M_s^2 d$, where $A$ is the exchange stiffness constant, $Q = \frac{2K_u}{M_s^2}$ the quality factor with $K_u$ magnetic anisotropy, $Ms$ the saturation magnetization, $D_i$ interfacial DMI and specimen thickness $d$. For the FGT with magnetic parameters[15] $A$ = ~1 pJ/m, $Q$ = ~16 and $Ms$ = 380 kA/m and the specimen thickness $d$ = 25 nm at which skyrmions and spirals coexist, we calculated the $D_i$ to be ~0.94 (±0.17) mJ/m$^2$. The measurement error bar was mainly associated with the measurement precision of specimen thickness. Our experimentally determined interfacial $D_i$ is well within the range of DMI values commonly reported in the heavy metal/Co-based multilayers[30].

To shed light on the origin of DMI and thus the observed chiral skyrmions and spirals in FGT nanolayers, we carried out density functional theory (DFT) calculations (details in the Methods) using the projected augmented wave method as implemented in the Vienna ab-initio simulation package (VASP)[33,34]. The self-consistent total energies for various noncollinear spin configurations in quintuple layers of FGT (Fig. 4) are evaluated with spin-orbit coupling considered using the constrained method implemented in VASP. Correspondingly, the DMI term can be estimated as demonstrated in bulk frustrated systems[35], insulating chiral-lattice magnets[36], and the interface systems[37,38].

The approach for three-sites DMI[19] calculation for a single hexagonal-lattice interface[37] between the magnetic ions and the mediated ions is now adapted for the Te/Fe interfaces in FGT. The quintuple-layered Te/Fe/FeGe/Fe/Te has two interfaces of Te/Fe and Fe/Te that are relevant for the DMI calculation. We first set the Te/Fe interface with clock wise (CW) and anti-clock wise (ACW) chiral spin configurations, while keeping the Fe spins in the other layers (FeGe and Fe/Te) oriented along the $z$-direction, as shown in Figure 4a and 4b. Similarly, in Fig. 3c and 4d the Fe/Te interface is set with the opposite chiral spin configurations. For the Te/Fe interface, the obtained $\Delta E_{DMI}$ is 38.7 meV or $d_1$ = 3.2 meV. In contrast, the $\Delta E_{DMI}$ for the Fe/Te interface is -39.2 meV or $d_1$ = -3.3 meV. It is noted that the positive (negative) value of $d_1$ means that the spins at the interface favor an ACW (CW)-type chiral configuration. Thus, the net $\Delta E_{DMI}$ for the Te/Fe/FeGe/Fe/Te quintuple layer is -0.5 meV, corresponding to 1.45 mJ/m$^2$ in the continuum limit, which is very comparable to the value of ~0.94 (±0.17) mJ/m$^2$ experimentally measured from thickness-dependent skyrmion-to-spiral transition.

Interestingly, the quintuple layer of Te/Fe/FeGe/Fe/Te in FGT has a mirror symmetry[13] with respect to the central FeGe atomic layer, see the spin configurations shown in Figs. 4a and 4d. In our calculations, such two spin configurations are of different DMI energies because the spin chirality has changed. On the other hand, comparing spin configurations in Figs. 4b and 4d with the same spin chirality, there is no other symmetry which would bring one into another, leading to again different DMI energies. That is, all four configurations (Figs. 4a-d) have different energies after imposing clock-wise or anticlock-wise spin chirality within one Fe layer. Here we emphasize that in the DMI calculations for either Te/Fe or Fe/Te interfaces, the spins in other Fe layers are fixed along $z$-axis (or $c$-axis); that is, a ferromagnetic ordering is preset, which is plausible scenario because of the strong out-of-plane magnetic anisotropy in FGT. This corresponds to the bulk or thick thin film cases. In short, the preset spin orders lead to the different DMI energies for Te/Fe and Fe/Te in the spin chirality calculations. Our first-principles calculations unveil a significant DMI in the quintuple layer, which explains the origin of the observed chiral spin textures in FGT nanoflakes in our experiments.

**Conclusion**

In summary, we have demonstrated the formation of chiral spin textures including spirals and skyrmions in a vdW ferromagnet FGT using Lorentz microscopy and off-axis electron holography. Systematic LM observations in a wide range of temperature and under applied field allow us to construct a detailed magnetic phase diagram for the chiral spin textures in FGT nanoflakes. We further obtained a specimen thickness-dependent transition between chiral spirals and skyrmions in FGT nanoflakes. The determined critical specimen thickness at which both spirals and skyrmions coexist together with other key magnetic parameters ($A$, $Ku$ and $Ms$) are used to estimate the interfacial DMI strength in FGT, yielding a value of close to 1 meV/m$^2$ that is comparable to the DMI values found in many heavy-metal/Co-based magnetic multilayers. Moreover, we employed first-principles calculations to elucidate the physical origin of chiral magnetism in FGT vdW magnet, and unveiled a sizable interfacial DMI at the Te/Fe and Fe/Te interfaces in the quintuple layer Te/Fe/FeGe/Fe/Te being responsible for the spin chiral selectivity in FGT. Importantly, the unique layered structure of FGT with magnetic $Fe_3Ge$ slab and nonmagnetic vdW Te-Te bilayer can be regarded as ideal magnetic multilayers with perfect atomic lattice via controlled synthesis. The found spin chirality in vdW ferromagnet $Fe_3GeTe_2$ is anticipated to motivate further studies on the fundamental spin physics and spin-based applications of vdW magnetic materials in reduced dimensions.

**Methods:**

Crystal growth

The growth of $Fe_3GeTe_2$ single crystals involves two steps. First, $Fe_3GeTe_2$ powders were obtained by solid-state reaction method. A mixture of powders of Fe (99.95 %), Ge (99.9999 %) and Te (99.999 %) with a stoichiometric ratio of 3:1:2 was placed in an alumina crucible and sealed in a quartz tube under vacuum. Such a reaction device was heated to 600 °C and kept for 5 days, followed by cooling to room temperature naturally. Second, $Fe_3GeTe_2$ single crystals were grown by chemical vapor transport method. A mixture of $Fe_3GeTe_2$ powder as precursor and iodine powder as vapor transport agent was placed in a 20-cm-long evacuated quartz tube, which was subjected to a temperature gradient of (700 - 650)/20 °C/cm for 10 days. Plate-shaped $Fe_3GeTe_2$ single crystals were collected in the tube regions of low temperature.

TEM characterizations

A mechanical exfoliation method with scotch tap was used to obtain FGT nanoflakes. The FGT thickness ranges from 10 to 100 nm, as measured by a log-ratio method in electron energy-filtered TEM (see measurement details in the Supplementary Materials). Two electron microscopes (JEM 2100F and ARM 200F, JEOL Inc., both operating at 200 kV) were used to characterize the crystallinity and microstructure of the FGT specimens.

LM and EH measurements

In situ LM observations were carried out in JEM 2100F microscope. Cooling holders (Gatan Inc.) of liquid-nitrogen type (model 626) and liquid-helium type (model 3100) were used, allowing specimen temperature to be varied from 350-100 K and 100-10 K, respectively. For LM imaging, the specimen was placed in magnetic field-free conditions (Lorentz mode) with the conventional objective lens turned off. The excitation of the objective lens was varied to apply magnetic fields normal to the specimen plane over a field range of between ± 5 mT (residual fields) and 2 T. EH experiments were carried out in ARM 200F microscope. For EH imaging, the biprism voltage was typically set to 140 V to produce an overlap interference width of 1.2 um and a holographic interference fringe spacing of 2.4 nm. For hologram recording, a cumulative acquisition approach was used to record 30 holograms (with an exposure time of 3 s for each hologram). Off-axis electron holograms were reconstructed numerically using a standard Fourier transform based method with sideband filtering using custom-designed Matlab codes.

First-principles calculations

The details of the density functional theory (DFT) calculations with the projected augmented wave method as implemented in the Vienna ab-initio simulation package (VASP) are given as follows. The Perdew-Burke-Ernzerhof (PBE) approximation[39,40] is adopted to describe the exchange-correlation interactions. The energy cutoff of 500 eV is set for the plane wave basis expansions. The *ab*-plane lattice constant[5] is set to be 3.99 Å. The k-point grids 12×3 are used to sample the Brillouin zones for the 1×4 supercell to guarantee good convergence. The self-consistent total energies for various noncollinear configurations (Fig. 4) are evaluated with spin-orbit coupling considered using the constrained method implemented in VASP.

**Acknowledgement**

This work was supported by the National Key Research and Development Program of China under Grant Nos. 2016YFA0300303, 2017YFA0303000, 2017YFA0504703, 2017YFA0302904 and 2017YFA0302901, the National Basic Research Program of China under Grant No. 2015CB921304, the National Natural Science Foundation of China under Grant Nos. 11774403, 11774391, 11774399 and 11804381, the Strategic Priority Research Program (B) of the Chinese Academy of Sciences under Grant Nos. XDB25000000, XDB07020000, the Scientific Instrument Developing Project of the Chinese Academy of Sciences under Grant No. ZDKYYQ20170002. Beijing Natural Science Foundation (Z180008).

## Author contributions

H.W., Z.-A.L. and H.F.T. performed Lorentz microscopy and electron holography experiments and analyzed magnetic imaging data. C.X.W. and Y.G.S. synthesized FGT single crystals. Y.Z. and H.B.Z. performed first-principles calculations. Z.-A.L. and J.Q.L. conceived and designed the studies. H.X.Y. and J.Q.L. supervised the project. All authors discussed the results and contributed to the manuscript preparation.

## Competing interests

The authors declare no competing interests.

## Additional information

Supplementary information is available for this paper at https://xxx

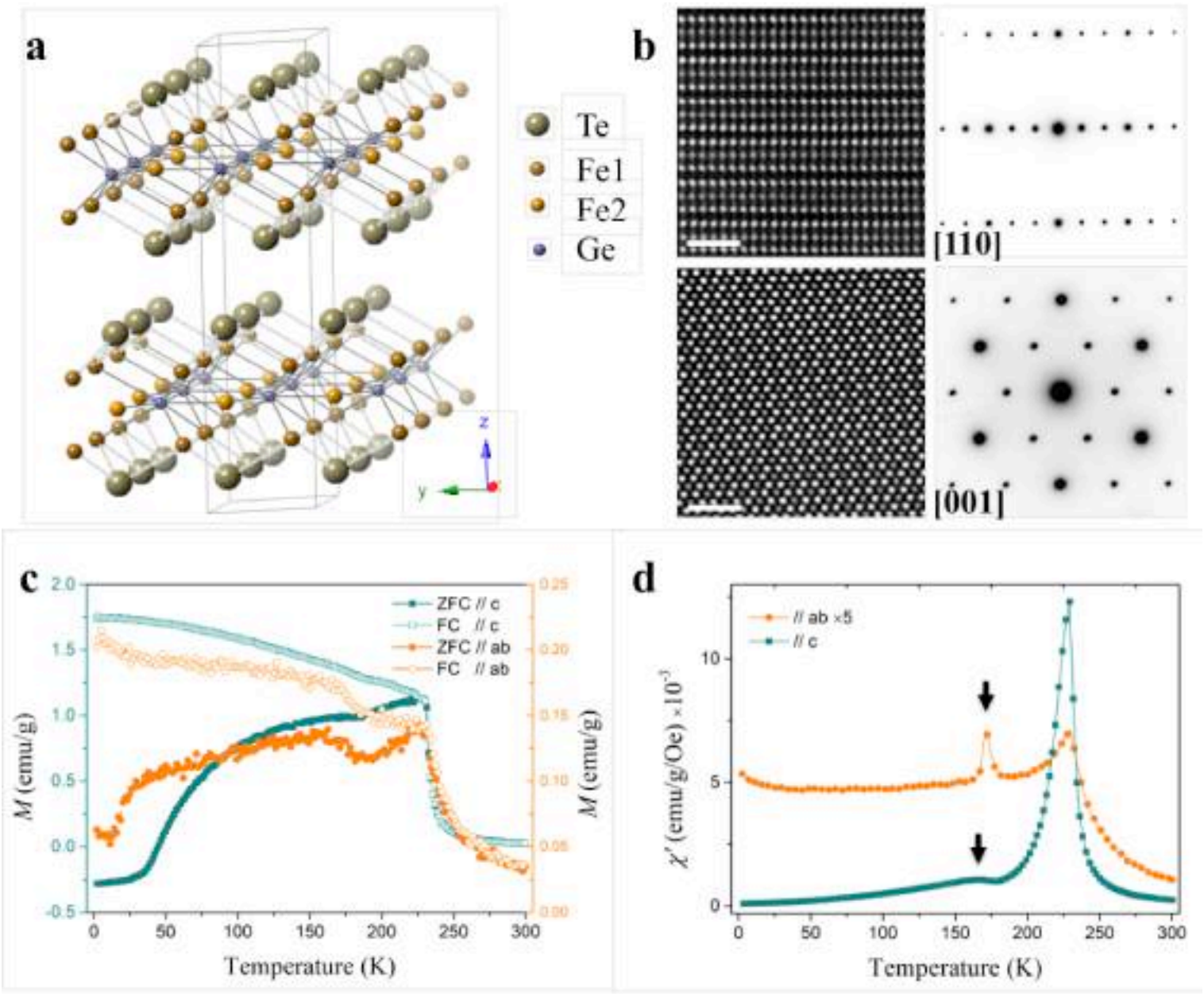

Fig. 1. Structural and magnetic properties of $Fe_3GeTe_2$ thin nanoflakes. (a) Atomic structure of $Fe_3GeTe_2$. (b) Atomic imaging and the corresponding electron diffraction patterns of [001] and [110]-oriented thin specimens. Scale bars correspond to 1 nm. (c) Temperature-dependent magnetization curves measured with 10-mT applied field along both *c*-axis and *ab*-plane. (d) AC magnetic susceptibility measured along *c*-axis and *ab*-plane.

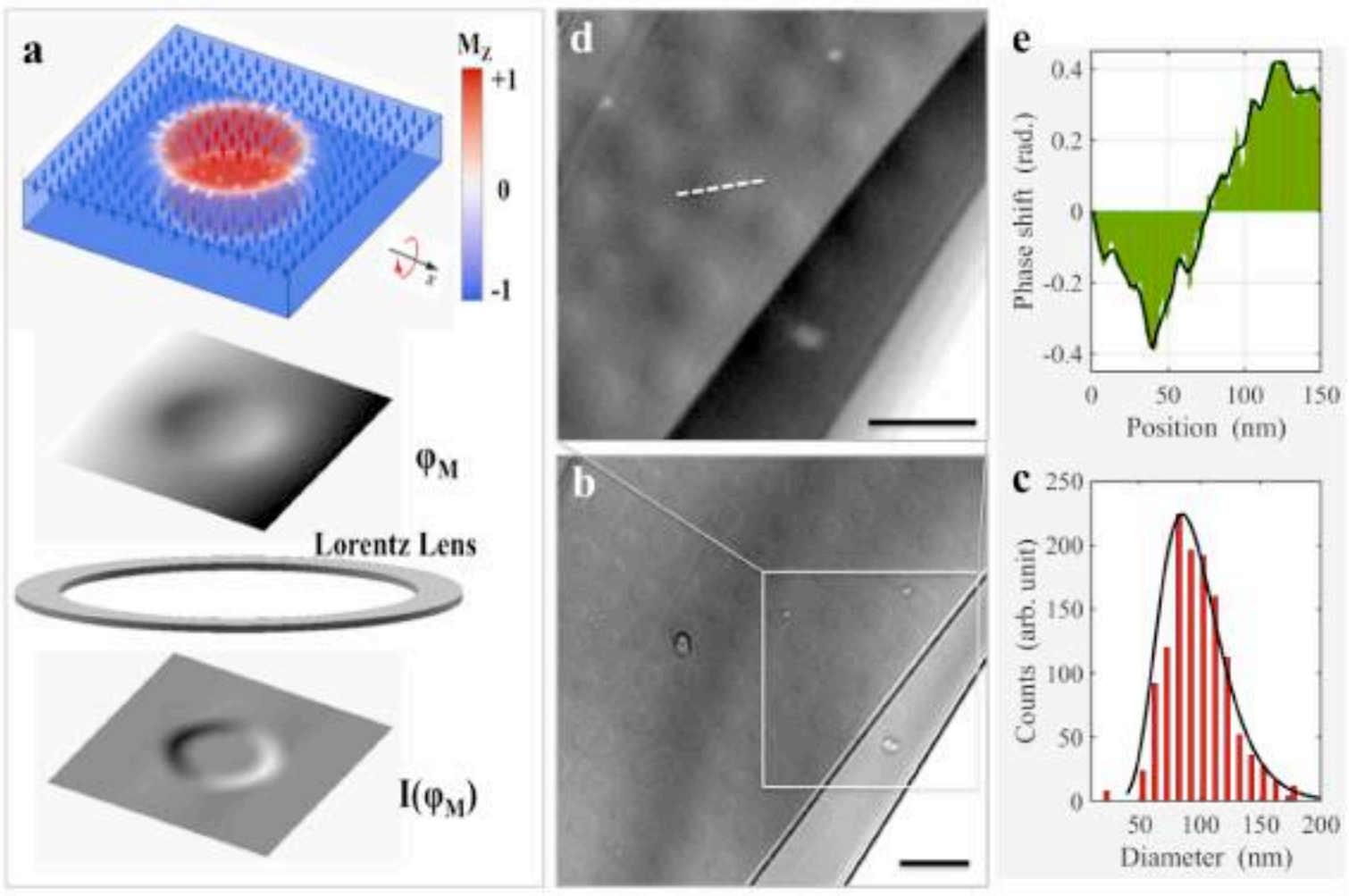

Fig. 2. Lorentz microscopy (LM) and electron holography measurements of Néel-type magnetic skyrmions at a temperature of 115 K and under zero-field. (a) Schematic representations of a single Néel skyrmion and its corresponding magnetic phase image and simulated Lorentz TEM images. (b) LM Fresnel image of $Fe_3GeTe_2$ nanoflake recorded with a defocus value of 600 um, displaying the formation of skyrmions. (c) Size distributions of Néel skyrmions determined from Lorentz images. (d) Total phase image measured from electron holography measurements. (e) Phase profile extracted from the line-scan region marked by dash-line in (d). Scale bars in (b, d) correspond to 200 nm.

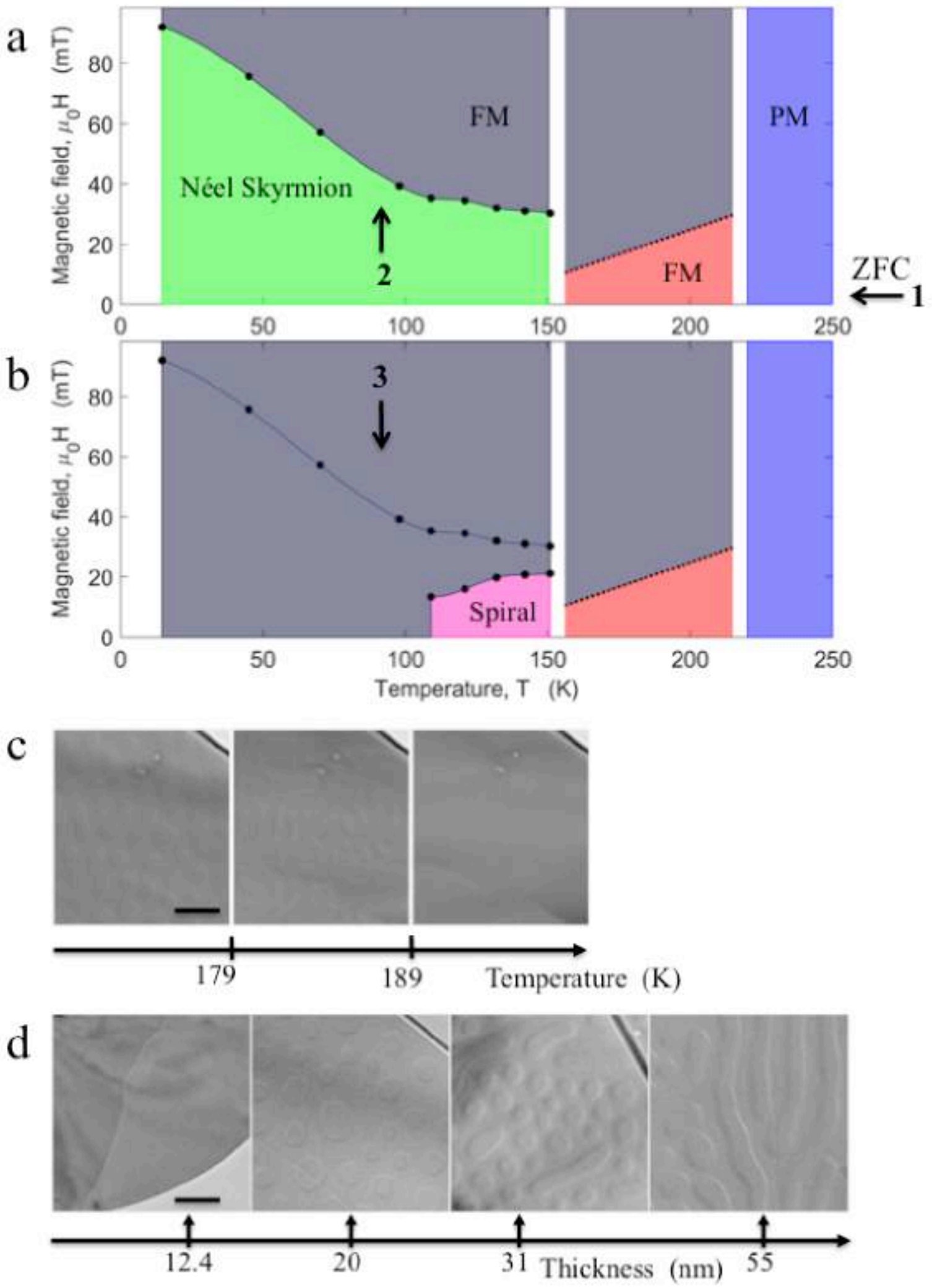

Fig. 3. Phase diagrams of magnetic structures in $Fe_3GeTe_2$ thin specimens. (a, b) Temperature and magnetic field (T-H) dependent magnetic states in a 20-nm-thick $Fe_3GeTe_2$ nanoflake. The sequences of varying specimen temperature and applying magnetic field are marked by arrows. The protocol of zero-field-cooling (ZFC) of the specimen from room temperature to 10 K was applied. (c) Upon heating under zero-field, magnetic transitions from skyrmions to stripes to paramagnetic states. (d) Thickness-dependent chiral spin textures in $Fe_3GeTe_2$ nanoflakes under zero-field conditions.

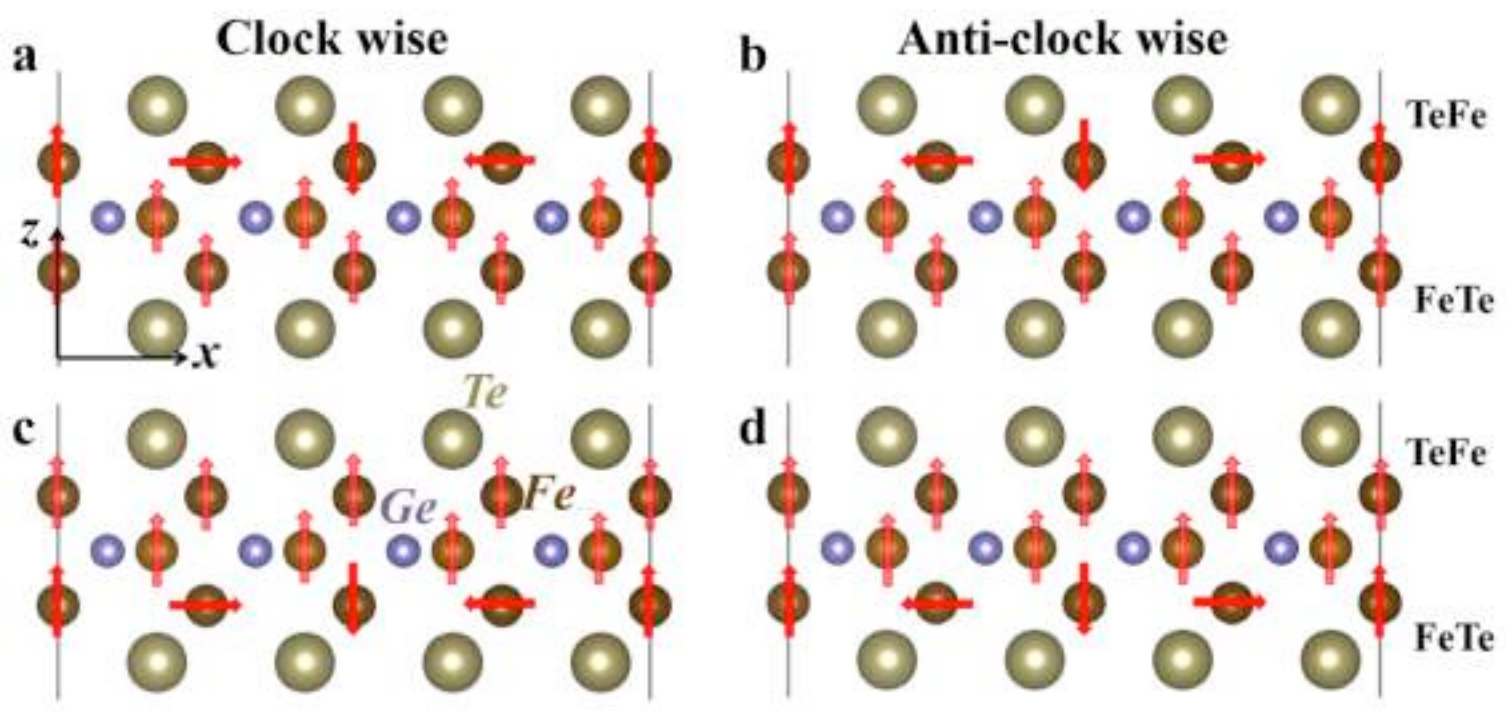

Fig. 4. (Color online) Chiral spin configurations in the Te-sandwiched $Fe_3Ge$ trilayer formulated as Te/Fe/FeGe/Fe/Te. (a) and (b) Side view (along crystallographic *a*-axis) of atomic structure of Te/$Fe_3Ge$/Te with clock wise (CW) and anticlock wise (ACW) spin configurations on Te/Fe, respectively. Spins of Fe atoms are represented by the arrows in 1x4 supercell. (c) and (d) are the same as (a) and (b) but for the Fe/Te interface.